\documentclass{IEEEtran}
\usepackage{cite}
\usepackage{amsmath,amssymb,amsfonts}
\usepackage{algorithmic}
\usepackage{graphicx}
\usepackage{textcomp}
\usepackage{colortbl}
\usepackage{xcolor}
\usepackage{threeparttable}
\usepackage{booktabs} % For better horizontal lines
\def\BibTeX{{\rm B\kern-.05em{\sc i\kern-.025em b}\kern-.08em
    T\kern-.1667em\lower.7ex\hbox{E}\kern-.125emX}}

\begin{document}

\title{A 200 dB Dynamic Range Radiation‑Hard Delta‑Sigma Current Digitizer for Beam Loss Monitoring}
\author{Luca Giangrande\thanks{
L. Giangrande is with CERN , Esplanade de Particules 1, Meyrin, Switzerland (e-mail: luca.giangrande@cern.ch) 
and with the University of Geneva, Department of nuclear and particle physics DPNC, Quai Ernest Ansermet 24, 1211 Geneva (e-mail: luca.giangrande@unige.ch).} }
\maketitle

\begin{abstract}
This manuscript describes a radiation-hardened current-mode delta-sigma ADC fabricated in a standard 130~nm CMOS technology and qualified for total ionizing doses up to 100~Mrad. The operational signal range achieved with 100~s integration window exceeds 200~dB. The converter is designed for beam loss monitoring applications in high-energy physics, where it must handle input currents spanning nine decades, from 1~mA down to 1~pA, while providing a fast 10~\textmu s response time for machine protection. To meet these conflicting requirements, the architecture exploits the inherent trade-off between resolution and acquisition time provided by delta-sigma conversion: a first-order architecture, sampling at 20~MHz, delivers 11~bit effective resolution within the critical 10~\textmu s window for critical currents around 1~mA. Integration times above 10~s enable the sub-picoampere resolution required for precise beam alignment and background monitoring. The chip integrates two independent channels, consumes 25~mW from a 1.2~V supply, and relies on radiation-hardening techniques such as triple-redundant digital logic, custom ESD protections and manual enclosed layout for critical analog transistors. Post-irradiation measurements up to 100~Mrad show no significant performance degradation, and the uncalibrated integral nonlinearity remains within [+4,--5] LSBs over the 1~mA to 5~\textmu A range. The converter's flexibility and radiation tolerance make it suitable not only for the HL-LHC beam loss monitoring upgrade but also for other precision current measurement applications in harsh environments.

\end{abstract}


\section{Introduction}
RADIATION-TOLERANT current digitizers primarily find their use in harsh environments where radiation monitoring is required, like in high energy physics experiments and nuclear power plants. For example, at CERN the radiation levels are carefully monitored around the Large Hadron Collider (LHC) in order to protect the machine components and to accurately position the absorber jaws of the mechanical beam collimation system. In both cases the sensing element is a gas filled ionization chamber \cite{BLM_IC1,BLM_IC2,BLM_IC3} that produces a fixed charge rate when subject to constant radiation levels; the corresponding electrical signal ranges from 1~mA to few picoamperes (pA) which represent respectively the maximum tolerable radiation energy, absorbed within the response time of the beam dumping system, and the typical background beam losses. Various current sensing interfaces exist either built with discrete components \cite{William} or as integrated circuits \cite{utopia} but their radiation tolerance limits impose a lower bound for the length of the cable connecting the interface with the sensor. Long cables limit the detection dynamic range because of to interference, noise charge associated with the parasitic capacitance and leakage currents. The radiation levels around the LHC are maximum at the collimators preceding the interaction points (IP3 and IP7) \cite{rad_lvl} and are expected to grow after the High Luminosity upgrade foreseen by 2025. The radiation around the collider originates from the losses of the beams circulating inside the machine, since the relevant quantity is the dose rate absorbed within one turn rather than its instantaneous value, the application band of interest spans from dc up to 12~kHz.   
An integrated radiation tolerant frontend has therefore been developed to comply with the stringent requirements of radiation monitoring in harsh environments. To this end a standard CMOS 130~nm technology was selected after being successfully characterized to tolerate a total ionizing dose (TID) of 200~Mrad \cite{Faccio, TID_130nm,Borghello} when adopting specific radiation hardness measures to prevent latch up, drain/source leakage and to tolerate single event upsets (SEU) and transients (SET). 
The remainder of this manuscript is organized as follows. Section II presents the system requirements and design drivers derived from the Beam Loss Monitoring (BLM) application and the HL-LHC upgrade constraints. Section III describes the architecture of the proposed delta-sigma current digitizer, discussing its resolution limit  and stability considerations while also providing the foundational concepts for readers less familiar with this architecture. Section IV details the circuit-level implementation, including the integrator, quantizer, adaptive clock control and DAC. Section V reports the experimental results, including noise performance, linearity, and dynamic range as a function of acquisition time, for samples irradiated with X-rays up to 100~Mrad. Finally, Section VI concludes the paper.

\section{System Requirements and Design Drivers}

\begin{table*}[h]
    \centering
    \begin{threeparttable}
    \caption{System Requirements and Design Drivers for the HL-LHC BLM Front-End ASIC}
    \label{tab:requirements}
    \begin{tabular}{p{3cm} p{3.3cm} p{9.2cm}}
        \toprule
        \textbf{Requirement} & \textbf{Value --- Range} & \textbf{Design Driver --- Origin} \\
        \midrule
        Input Current Range & 1~pA to 1~mA \newline (180 dB) & Ionization chamber physics: dynamic range spans from minimum detectable losses for beam alignment to full-scale quench-inducing events. \\
        \midrule
        Response time for \newline currents $\geq 0.5 mA$ & 10~µs & Beam dump response time: the system must react within 10~µs to trigger a beam bump and prevent sub-systems damage and magnet quenching. \\
        \midrule
        Radiation Tolerance \newline (tested survival) & $\geq 100$~Mrad \newline (1 MGy) & HL-LHC radiation level projections: electronics must survive the integrated dose expected over the upgraded accelerator lifetime. \\
        \midrule
        Technology Node & CMOS 130~nm & CERN-qualified process for 200~Mrad tolerance; offers optimal balance between radiation hardness, analog performance, density, and power. \\
        \midrule
        Supply Voltage & 1.2~V (thick-oxide devices\newline compatible with 2.5~V) & Standard for 130~nm CMOS; trade-off between power consumption, signal swing, and noise performance. \\
        \midrule
        Number of Channels \newline per Chip & 2 analog readout channels & System density: approximately 4000 ionization chambers served by 750 acquisition modules, resulting in 3 to 4 ASICs per module. \\
        \midrule
        Cable Length \newline Compatibility & Coaxial up to 50 m  & Tunnel infrastructure: long connections between chambers and electronics introduce attenuation and filtering. Proximity to high radiation accelerates aging. \\
        \midrule
        Communication \newline Protocol & LpGBT \newline (e-Link compatible) & Compatibility with HL-LHC data aggregation and timing distribution infrastructure. \\
        \midrule
        Communication \newline Redundancy & Dual communication \newline channels & System reliability: ensure continuous operation and data transmission even in case of link failure. \\
        \midrule
        Radiation-Induced \newline Errors & SEU resilience \newline (TRM \tnote{a} digital design) & High-energy hadron fluxes expected in HL-LHC tunnel; mitigation of single-event upsets is mandatory. \\
        \midrule
        EMI Immunity & Robust against \newline electromagnetic interference & Harsh accelerator environment with high-power RF systems, switching transients and non-negligible magnetic fields. \\
        \midrule
        Out-of-Range Input \newline Handling & Operational for currents \newline beyond nominal range & Ensures collider protection: system remains responsive during extreme events and recovers without latch-up or damage. \\
        \midrule
        Chip Physical \newline Dimensions & $\leq 4 \times 4~mm^2$ & Fits within 64-pin QFP cavity; balances functionality, power dissipation, and mechanical robustness. \\
        \bottomrule
    \end{tabular}
    \begin{tablenotes}
            \item[a] Triple redundant modular design with majority voting and spatial spread.
        \end{tablenotes}
    \end{threeparttable}
\end{table*}

The BLM system \cite{BLM_SYS} is a critical protection element of the LHC accelerator complex, designed to prevent quenches of superconducting magnets and equipment damage caused by excessive radiation following beam misalignment. The system continuously measures radiation levels along the accelerator using approximately 4000 ionization chambers installed throughout the 27~km of the circular collider. When ionizing radiation passes through the gas-filled chambers, the resulting charge is extracted thanks to a high-voltage bias (1.5~kV) and converted into a current proportional to the instantaneous beam loss. In the original LHC design these signals are transmitted via coaxial cables, up to 800~m in length, to about 750 acquisition modules for processing and analysis.

To ensure compatibility with the higher luminosity and correspondingly increased radiation levels expected from the HL-LHC upgrade, a new generation of front-end electronics must meet stringent performance, reliability, and environmental constraints. The enhanced radiation tolerance enables the readout electronics to be installed closer to the sensors, thereby improving signal integrity. However, depending on the radiation levels and associated aging effects, the electronics must also remain compatible with cable lengths of up to 50~m to preserve installation flexibility, although with relaxed requirements for the non-critical current levels ($\leq 0.5$~mA).

Another critical requirement is the response time for triggering an emergency beam dump. The beam circulation time inside the machine is $89 ~\mu s$, and the electronics must be capable of initiating a dump within $10~\mu s$ of detecting a fault condition.
Table~\ref{tab:requirements} summarizes the key system-level requirements and the design drivers that shape the ASIC architecture.

The stringent requirements of a nine-decade dynamic range and a $10~\mu\mathrm{s}$ response time would imply impractically high instantaneous resolution for a direct conversion approach. However, this requirement can be relaxed when considering the limited signal bandwidth of $12~\mathrm{kHz}$ and the variable acquisition time, which can extend up to tens or even hundreds of seconds for less critical currents. This long-acquisition, high-resolution regime corresponds to beam alignment tracking, where the signals can be considered essentially static.
Consequently, the first-order current-mode delta-sigma ADC emerges as an ideal candidate for beam loss monitoring applications. While higher-order modulators offer superior noise shaping and can achieve better resolution for a given oversampling ratio, their performance advantage diminishes at lower frequencies, where the effective resolution is ultimately limited by the DC gain of the loop filter, namely a parameter that is independent on the converter's order. Moreover, their stability is harder to achieve and less robust to perturbations, particularly under saturation or overload conditions, where they are prone to prolonged recovery transients or even deadlock. In contrast, the first-order architecture provides predictable, unconditional stability and graceful saturation recovery, making it ideally suited for a machine protection system where reliability and deterministic behavior are paramount.

The extreme radiation requirement, exceeding 100~Mrad total integrated dose (TID), dictates both technology choice (130~nm CMOS) and design methodology. Digital circuits must be triplicated with majority voting to mitigate SEUs, while analog blocks require custom enclosed layout and more stringent latch-up constraints to preserve functionality and performance after cumulative dose degradation.

The use of long cable, extending up to 50~m, adds substantial capacitance to the ionization chamber's intrinsic 350~pF, bringing the total load seen by the front-end to 5~nF. This extended capacitance demands a low input impedance to mitigate cable-induced time constants, while also requiring the circuit to withstand the noise and interference inevitably picked up over such transmission lengths.

Finally, the system's role as a machine protection element imposes absolute reliability requirements. Redundant communication channels, SET resilience, and graceful out-of-range behavior are not optional but mandatory to ensure the BLM system can fulfill its critical function throughout the HL-LHC operational lifetime.

\section{Converter architecture}
\begin{figure}
    \centering
    \includegraphics[width=0.45\linewidth]{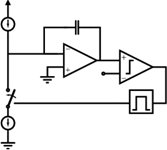}
    \caption{Block diagram of the first-order delta-sigma ADC. The modulator operates in current mode, with both the sensor input and the feedback compensation implemented as currents. The clocked comparator outputs a digital bitstream, which is subsequently processed and also fed back to control the compensation current.}
    \label{fig:mod_block_diag}
\end{figure}

The core of the measurement system is a continuous-time, first-order, single-bit delta-sigma modulator, which converts the sensor’s input current into a digital bitstream. A simplified schematic of this ADC is shown in Figure~\ref{fig:mod_block_diag}.

The front-end of the modulator consists of an active integrator, which serves two critical purposes: it collects the ionization charge generated by the sensor over time and acts as a filter for the feedback loop. 
The integrator’s output is continuously monitored by a comparator, clocked at a default frequency of 20 MHz derived by dividing the 80 MHz clock used for the digital circuitry. This frequency represents the optimal trade-off between the integrator’s bandwidth and the analog-to-digital conversion range achievable within the $10~\mu$s response time.
The integrator output is compared to a reference threshold at the clock's rising edges, effecting both sampling and 1~bit quantization. The comparison result drives a time-varying negative feedback correction that prevents the integrator from saturating. 
This feedback is implemented via a 1~bit digital-to-analog converter (DAC), consisting of a current source and a switch controlled by the comparator's output. The precision of the compensation charge depends on the stability of the reference current and the clock period.
The effectiveness of the sigma-delta conversion relies on the continuous accumulation of the net difference between the input and feedback signals, and on its attenuation by the loop filter gain.
A high DC loop gain guarantees that the time-averaged compensation signal closely follows the average input current. To first order, therefore, the achievable resolution of the converter is proportional to the DC gain of the integrator.

The bitstream produced by the quantizer is the digital representation $D_\textit{out}$ of the compensation signal whose average value is directly encoded in the density of '1's within this bitstream. 
For static inputs, therefore, the upper bound under ideal 1~bit averaging of the achievable resolution corresponds to the number of distinguishable codes, expressed as:

\begin{equation}
    Resolution=log_2\left(\frac{D_{out, max}}{D_{out, min}}\right)=log_2\left( f_{samp} \cdot T_{acq}\right)
    \label{eq:resolution}
\end{equation}
where $f_{samp}$ is the sampling frequency and $T_{acq}$ is the acquisition time per measurement. 

\begin{figure}
    \centering
    \includegraphics[width=0.9\linewidth]{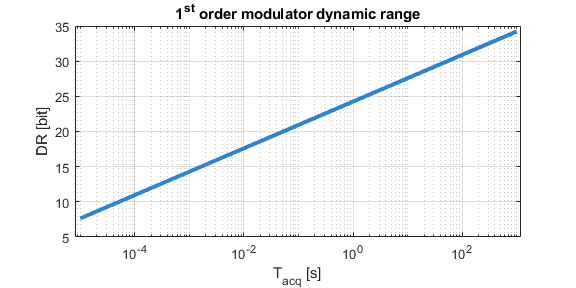}
    \caption{Calculated intrinsic dynamic range of a first-order delta-sigma modulator as a function of acquisition time, for static input signals. The dynamic range spans from a single high pulse ('1') to an all-ones bitstream. The plot assumes a constant sampling frequency of 20 MHz.}
    \label{fig:Dyn_rang}
\end{figure}

Eq.~\ref{eq:resolution} highlights the intrinsic trade-off between signal bandwidth and conversion resolution characteristic of the sigma-delta topology, making it ideally suited for accurate, high-resolution measurements of slowly varying signals. The bandwidth limitation comes from using longer bitstreams to evaluate the instantaneous average. The relationship is depicted in Figure~\ref{fig:Dyn_rang}

\subsection{Decimation filtering}
\begin{figure}
    \centering
    \includegraphics[width=\linewidth]{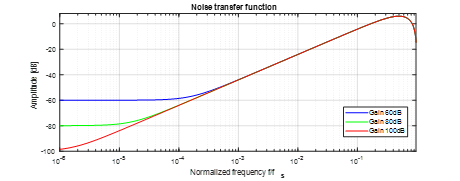}
    \caption{The noise transfer function (NTF) illustrates how signals injected after the integrator, acting as a loop filter, are shaped when referred to the input. The complementary filter response provides substantial attenuation in the low-frequency band of interest, balanced by a contained amplification at higher frequencies.}
    \label{fig:NTF}
\end{figure}

The raw bitstream retains the large quantization error inherent to single-bit quantization, which encodes only the sign of the integrator output relative to a threshold. Under the action of the loop-filter, however, this error is attenuated at low frequencies and it's energy content is shaped  toward the higher end of the spectrum. Figure~\ref{fig:NTF} illustrates the effect of the integrator's finite gain-bandwidth product on the modulator's noise-shaping characteristic.

In this regime, the excess quantization error can be removed through real-time filtering of the bitstream. The industry standard is an n-th order sinc filter whose transfer function is: 

\begin{equation}
H_{M,n}(z) = \left[ \frac{1}{M}\frac{1-z^{-M}}{1 - z^{-1}} \right]^{n}
\label{eq:sinc_filter}
\end{equation}
where M is the bitstream length and n the number of cascaded first-order filters. The case n=1 corresponds to a simple moving average, representing the sweet spot for low latency and hardware efficiency, while n=3 provides optimal accuracy. This low-pass filtering also enables the symbol rate to be reduced accordingly without any loss of information.

%H(z) = \left[ \frac{z^{-1}}{2 \cdot \left(1 - \frac{z^{-1}}{2}\right)} \right]^{n}= \frac{z^{-n}}{2^{n}} \cdot \left[ \sum_{k=1}^{n} {n \choose k} \left( \frac{-z^{-1}}{2} \right)^{k} \right]^{-1}

The inherent trade-off between accuracy and bandwidth offered by sigma-delta conversion can be further exploited by making the outputs of these multiple filter stages available simultaneously. Thereby allowing the BLM system to concurrently use measurements with varying bandwidth and resolution for optimal beam alignment and collimation. 

\subsection{Non-idealities} Beyond the usual limitations of analog components such as noise, offset, slew rate, finite gain-bandwidth, quantizer delay, and switch non-idealities, the converter architecture itself introduces notable deficiencies.

One such phenomenon is the appearance of limit cycles due to the static input signal being a rational multiple of the compensation current. Under these conditions, the bitstream exhibits periodic regularity, concentrating quantization error energy at specific frequencies in the output spectrum. However, this effect is not critical: as zero-mean tones, their primary impact, if not properly mitigated, is a potential 1-LSB oscillation between consecutive output samples.

A second issue arises from transitory perturbations, such as SEUs or out-of-range inputs, that drive the integrator into saturation. During saturation, the bitstream locks to a constant sequence of '1's, causing a temporary spill of quantization error into lower frequencies until recovery is complete and normal operation resumes. However, this converter overload does not prevent the detection of critical radiation levels and the initiation of an emergency beam dump. Moreover, unless an ESD protection discharge occurs, the charge produced by the ionization chamber is conserved and its average rate is correctly measured by the converter.

\subsection{1-st order modulator stability} 
A first-order delta-sigma modulator offers an unconventional perspective on stability. The loop contains an ideal discrete-time integrator with a pole at $z=1$, which is marginally stable: in the absence of feedback, a step input produces an output that grows unbounded, albeit linearly rather than exponentially. The boundedness of the modulator is provided by the action of the nonlinear feedback implemented through the 1~bit quantizer and DAC.

This nonlinear feedback forces polarity reversals whenever the integrator state crosses the decision threshold, preventing monotonic growth of the state variable. For input signals within the available feedback range, the system does not converge to a static equilibrium but instead evolves toward a bounded oscillatory trajectory in state-space. In this sense, the modulator behaves similarly to a relaxation oscillator whose amplitude is constrained by feedback.

From a purely linear viewpoint, the sustained oscillatory behavior may resemble that of a feedback system with insufficient phase margin. However, the underlying mechanism is fundamentally different: the oscillation is not caused by linear pole placement but by the discontinuous switching action of the quantizer. It is therefore intrinsic to the intended operation of the converter rather than a symptom of linear instability.

The staircase-like nonlinearity of the quantizer and DAC introduces irregular switching patterns for most input values. This irregularity is the mechanism that spreads quantization error energy toward higher frequencies, enabling noise shaping and preventing its concentration into discrete tones, provided pathological limit cycles are avoided.

Despite its nonlinear nature, the modulator can be approximated as quasi–linear time-invariant for analytical purposes. By linearizing the quantizer and DAC around their average operating point, classical feedback theory can be applied as a first-order approximation, yielding useful insight into loop gain, signal and noise transfer functions, and stability margins within the intended operating range.
When the input exceeds the available feedback current, the state escapes this bounded trajectory and grows until physical saturation occurs; normal bounded operation resumes once the input returns within its operational range.

%System level diagram with additional moving average LPFs.

\begin{figure}
    \centering
    \includegraphics[width=0.9\linewidth]{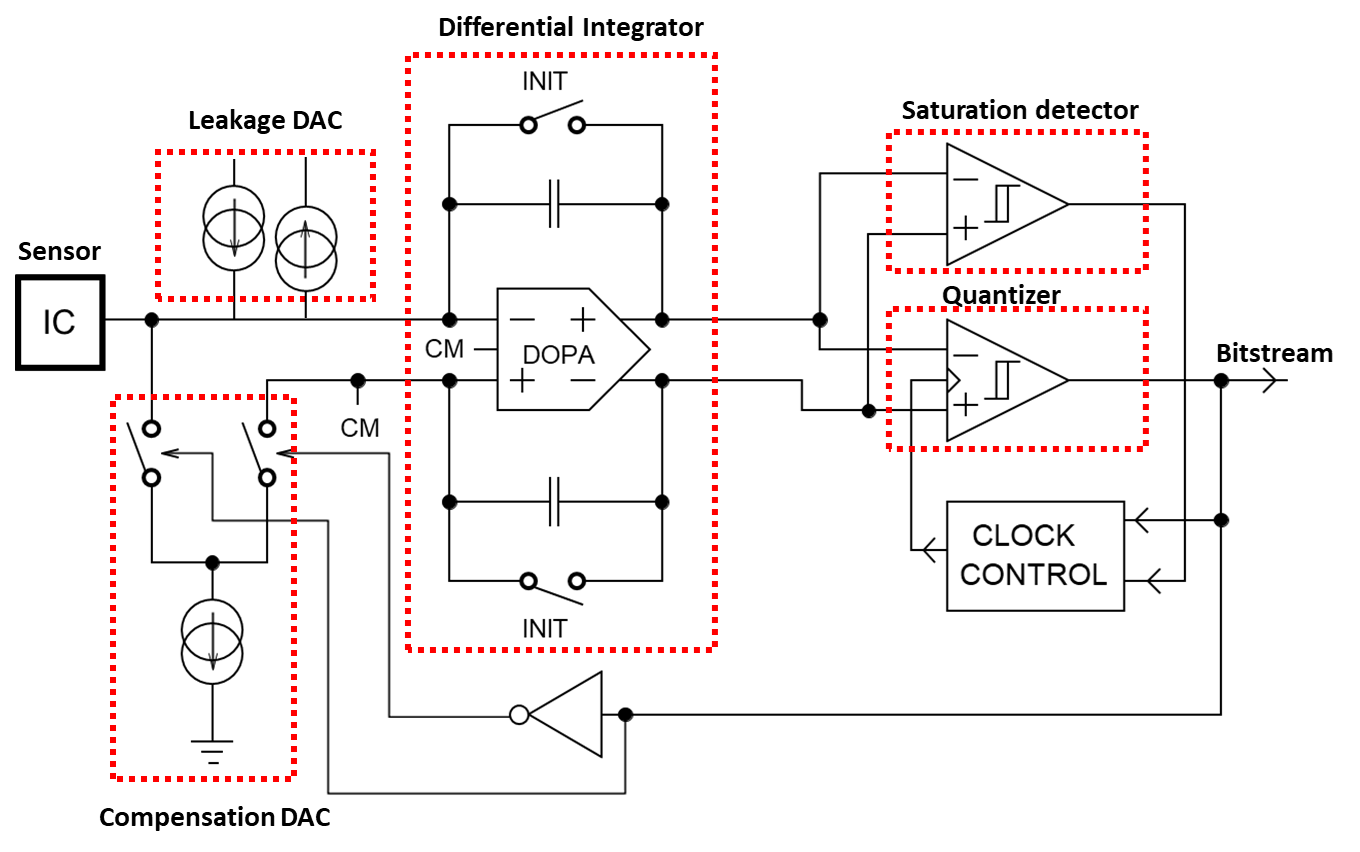}
    \caption{Simplified schematic of the implemented ADC. The fully differential integrator converts the single-ended sensor current into a differential signal. Adaptive clock control, complementary switching for the reference current, and reset switches ensuring DC loop continuity are highlighted. Integrating capacitors are sized to 60~pF.}
    \label{fig:diff_mod_block_diag}
\end{figure}

\section{Circuit implementation}
Building upon the operational principles of the delta-sigma modulator, this section delves into the circuit-level realization of the converter. 
%The implementation is presented hierarchically, beginning with the high-level architectural choices that shape the design constraints, and progressively descending into the transistor-level details of the key building blocks. 
This approach highlights how system-level considerations such as noise performance, power consumption, and bandwidth are translated into concrete circuit topologies and device-level decisions.

% Explain the se 2 diff conversion by means of the CMFB which is then designed with the same bandwidth
The main architectural features of the analog-to-digital converter implemented on silicon are shown in Figure~\ref{fig:mod_block_diag}. The integrator employs a fully differential topology, which converts the single-ended, unipolar signal from the sensor into a differential representation. This conversion results from the interaction between the integration path, which charges exclusively the upper output node, and the common mode regulation, which induces a complementary current in the lower integration capacitor to stabilize the output common mode. 
This approach provides an increased output swing and enhanced immunity to on-chip digital interference. 
To optimize power consumption and accuracy, an adaptive clock control scheme is implemented, scaling the operating frequency for smaller input signals. The compensation current source is steered using complementary switches, minimizing transient settling errors and improving overall accuracy. 
%Additionally, reset switches with a finite off-resistance ensure loop continuity even under DC conditions, preserving the integrity of the feedback path.    
% Reset switches in T-configuration to minimze uncontrolled leakage current towards the input.

\subsection{Amplifier}
At the heart of the fully differential integrator lies the operational transconductance amplifier (OTA), the most critical analog block in the signal chain. Its performance directly governs the overall linearity, noise, and settling behavior of the modulator, ultimately determining the achievable resolution of the converter.

\begin{figure}
    \centering
    \includegraphics[width=1\linewidth]{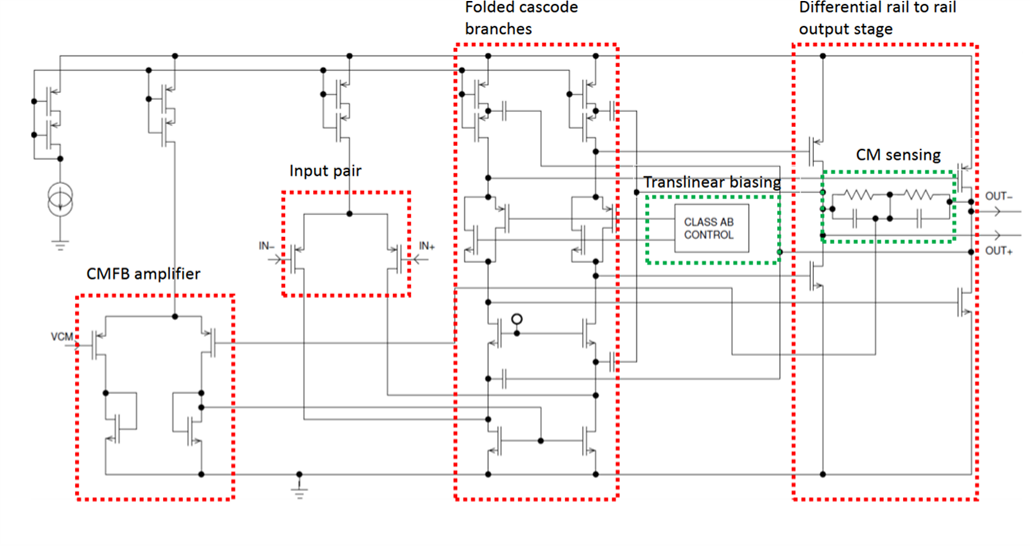}
    \caption{Simplified schematic of the operational transconductance amplifier (OTA). A two-stage, cascode-compensated topology with a PMOS folded cascode input stage and a class-AB rail-to-rail output stage biased through a translinear loop. Common-mode feedback (CMFB) regulation amplifier shown in the bottom left.}
    \label{fig:OAmp_schematic}
\end{figure}

The amplifier, shown in Figure~\ref{fig:OAmp_schematic}, employs a fully differential, two-stage, cascode-compensated topology with a rail-to-rail output stage capable of class-AB operation. The first stage is implemented as a folded cascode using a PMOS input pair, which provides high gain and good common-mode rejection. The choice of PMOS devices for the input pair is motivated by their lower susceptibility to radiation-induced flicker noise increase, as their buried conduction channel is less affected by damage at the interface with the gate oxide.

The differential output is realized with a push-pull configuration, biased through a translinear loop to precisely control its quiescent current. The class-AB configuration prevents slew-rate limitations during large-signal excursions without incurring excessive power penalty. 
%The class-AB regime is obtained through the dynamic control of the push-pull transistors bias point

The design is adapted from \cite{Hogervorst} and optimized for the integrator requirements. It achieves an open-loop gain exceeding 95~dB, a unity-gain bandwidth above 50~MHz, and a phase margin of 74~degrees. The input-referred offset is 0.62~mV, while the total current consumption remains below 4~mA from a 1.2~ V supply. The thermal output noise level is 3.6~nV/$\sqrt{Hz}$ with a flicker corner frequency around 20~kHz. The amplifier's high DC gain ensures effective attenuation of quantization noise at low frequencies, while its unity-gain bandwidth strikes an ideal compromise between settling speed and noise performance. 

%% Actually there isn't a closed form for Q_n(A_dc) the following analysis does not account for noise shaping and "in-band only" noise.
%Given an input capacitance dominated by the ionization chamber itself ($C_{sens}=350~pF$), the open-loop gain required to achieve a DC quantization error attenuation consistent with 15-bit resolution is approximately:
%\begin{equation}
%    A_v = \frac{\mathrm{FS}}{2\sqrt{12}}\cdot\frac{C_{int}}{C_{sens}+C_{int}}\cdot\frac{2^{15}}{\mathrm{FS}}\approx 90~dB
%\end{equation}
%where $\mathrm{FS}$ is the converter's full scale and $C_{int}$ is the integration capacitance. 

\subsection{Quantizer} 
\begin{figure}
    \centering
    \includegraphics[width=0.5\linewidth]{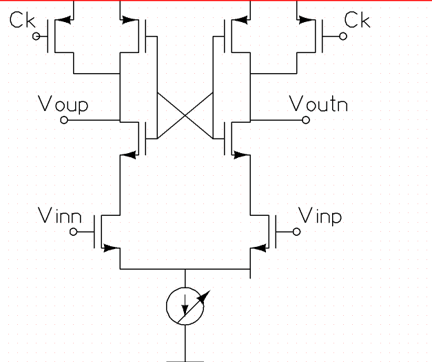}
    \caption{Simplified schematic of the synchronous comparator. A strong-arm latch topology with tunable bias current (3~bit, $160~–~640~\mu$A) and selectable triple-redundant configuration for SEU mitigation (not shown).}
    \label{fig:clock_comp}
\end{figure}

The 1~bit quantizer, Figure~\ref{fig:clock_comp}, is implemented as a synchronous comparator using a strong-arm latch topology, which is reset during the low phase of the clock and released on its rising edge. The total current drawn from the supply—before the comparator is fully steered—is limited by a tunable bias current programmable from 160 to $640 ~\mu A$ with 3~bit resolution. This bias current also influences the comparison speed and noise performance.

The comparator output generates a control signal for the compensation current, which remains active for a full clock cycle—from the current rising edge until the next. This approach optimizes the correction current sizing by relying on the well-controlled clock period rather than its duty cycle.

To enhance the system's radiation hardness, a selectable triple-redundant comparator configuration can be enabled. In this mode, three parallel comparators are activated, and a majority voting logic circuit produces the final output.

The comparator achieves, in the minimum current configuration, an input-referred offset of $90 ~\mu V$, an input-referred noise of $85 ~\mu V_{rms}$, and exhibits a transient voltage kickback of 175 mV.

\subsection{Adaptive clock control}

The dynamic clock rate control is a feature designed to minimize switching activity in the proximity of the analog circuitry when small currents are being measured. The clock control logic monitors the output bitstream and adjusts the clock frequency—ranging from 20 MHz down to 38 Hz—based on the number of consecutive identical symbols detected. The on-pulse duration remains fixed, as it determines the value of the compensation charge together with the reference current. The clock rate is doubled or halved depending on which string of identical symbols is observed.

The modulator's optimal operating mode is found in the neighborhood of the middle of the input range, where the commutation rate is maximized and the quantization error is uniformly spread across the spectrum. The length of consecutive symbols required to trigger a clock rate change drives a trade-off between the response time of the additional control loop, the target switching activity, and the interaction with the signal dynamics. This threshold is programmable and is typically set to three consecutive symbols after allowing sufficient time for the modulator to settle.

For a given static current, the minimum operating frequency is determined by the value that would cause the integrator to saturate within two clock cycles.

To mitigate the long settling time following a large input step, an asynchronous comparator is introduced. It monitors the integrator output and resets the clock to its maximum frequency if saturation is detected, thereby overriding the dynamic control loop and ensuring a fast response on the order of tens of microseconds.

\subsubsection{Asynchronous comparator}
\begin{figure}
    \centering
    \includegraphics[width=0.8\linewidth]{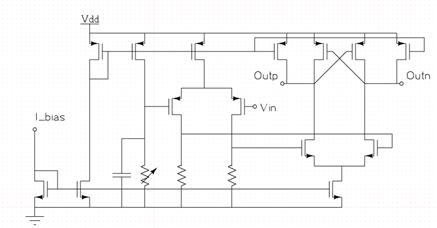}
    \caption{Asynchronous comparator for saturation detection. Two-stage open-loop amplifier with resistive-load preamplifier for kick-back attenuation. Output stage with regenerative cross-coupled load for faster commutation. The threshold (10–300 mV) is generated by a bandgap-derived current through a tunable resistor, with a 68~pf capacitor for noise filtering.}
    \label{fig:asyn_comp}
\end{figure}

The asynchronous comparator, Figure~\ref{fig:asyn_comp}, is implemented as a two-stage open-loop amplifier with a total gain of 43 dB. The first stage is a low-gain preamplifier designed to reduce kickback toward the integrator; it employs a common-source topology with resistive loads. The second stage features a regenerative cross-coupled load that guarantees rapid decision-making through positive feedback. Diode-connected loads prevent saturation, thereby ensuring a faster recovery.

The comparator monitors only the negative integrator output and compares it to an internal threshold voltage. This threshold is generated by steering a bandgap-derived current of $20 ~\mu A$ through an adjustable resistor producing reference voltages from 10 to 300 mV. 
The comparator exhibits an input-referred offset of 2.2 mV and a hysteresis of 7 mV provided by the positive feedback. Its current consumption is $120 ~\mu A$.

% Make a better picture with more details. Logic, bias, decoupling, etc.

\subsection{DAC}
The feedback signal is generated by a 1~bit DAC, which can be trimmed using a 5~bit control word to adjust the converter's full-scale range according to the expected input current, spanning from $20 ~\mu A$ to $1.06~mA$. In order to prevent the modulator from saturating, the feedback current must be at least one LSB larger than the maximum input current.

The total output-referred noise of the DAC is $530~\mathrm{nA_{rms}}$.Relating this to its nominal output and expressing the ratio in bits yields:  

\begin{equation}
 Resolution_{DAC}=\log_2\left(\frac{I_{\mathrm{max}}}{I_{\mathrm{noise,rms}}}\right)\approx10.8 ~bits
\end{equation}
This contribution limits only the conversion of full-scale signals, as it is not injected at every sampling period for lower currents. Consequently, it does not impair accuracy in the mid and lower signal ranges. Since the modulator scales a fixed compensation current with the input, the associated noise scales proportionally. As a result, the DAC's effective noise diminishes with decreasing input current and is further reduced by bitstream averaging.

% calculate it as charge and account for the standard error \sqrt{N_{bit1}.
The DAC is implemented as a current mirror using NMOS devices, incorporating cascode transistors and degeneration resistors to enhance noise performance. To prevent radiation-induced mismatch, unused devices are kept active by continuously drawing current from the common-mode voltage (Vcm), rather than being switched off.

\begin{figure}
    \centering
    \includegraphics[width=0.9\linewidth]{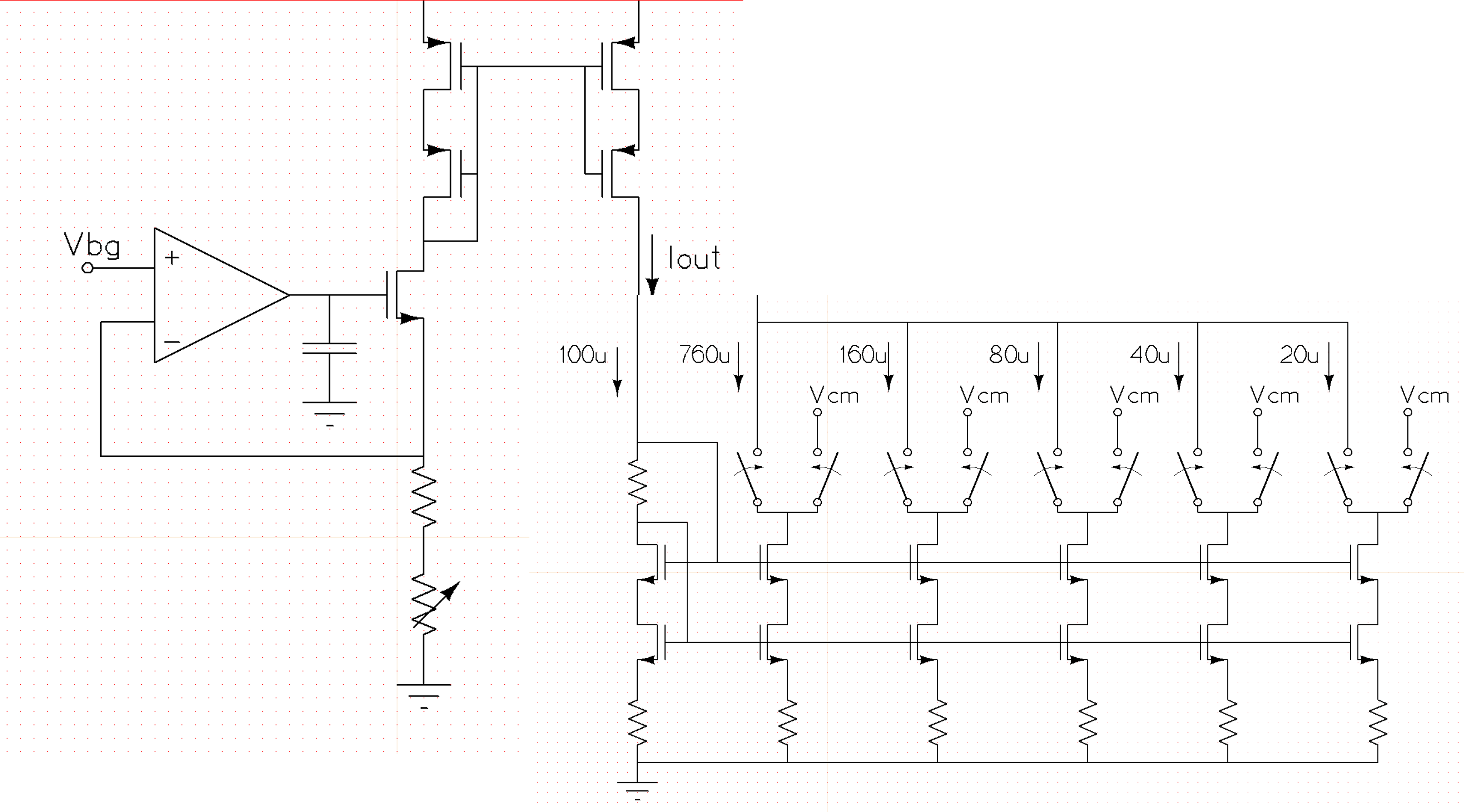}
    \caption{Simplified schematic of the 1~bit feedback DAC. A current mirror with cascode transistors and degeneration resistors, trimmed via a 5~bit control word to adjust the full-scale range from $20~\mu$A to 1.06 mA. Unused devices are kept active to prevent radiation-induced mismatch.}
    \label{fig:DAC_schematic}
\end{figure}

%\subsection{Radiation hardness techniques}
%P-moses are less sensitive to radiation induced defects because of their buried conduction channel. Veto. Leakage (Vcm- Vdd).

%\section{Calibration}
%Calibration can be performed regularly by exploiting the known radiation level decay rate in absence of circulating beam.  Range adjustment. Leakage compensation and tracking with auxiliary channel.Leakage compensation is particularly challenging in this context, as the currents involved are in the picoampere range and are often not accurately modeled. To avoid relying on external components or impractical current mirroring schemes, an optimal solution is to exploit the gate leakage current inherent in transistors with gate oxide thicknesses on the order of a few nanometers. This approach leverages the fact that the voltage drop across the equivalent resistance remains stable when the common-mode reference is derived through ratiometric division from a higher-voltage source; preferably a bandgap reference, or alternatively a filtered supply voltage. By periodically compensating for this leakage through a dedicated calibration step, charge is prevented from being stolen from the sensor, thereby improving conversion accuracy and long-term stability. Moreover this approach is quite efficient in terms of area as drawn devices do not need to be considerably larger than the input pair of the integrator's amplifier. 
%Finally the radiation induced drift is not large and it tracks the 

\section{Measurements Results}
The ASICs were irradiated without their package lids, using CERN's in-house 50 kV, 3 kW X-ray generator (SEIFERT RP149). The total accumulated dose reached 100~Mrad, with an absorbed dose rate of approximately 8.82~Mrad/h referred to $SiO_2$. During irradiation, the chips were operated under worst-case conditions: a supply voltage of 1.575 V (approximately 30\% above nominal), a sampling frequency of 20~MHz, and a static input current of 1~mA, while the operating temperature was maintained at 25~°C.

Post-irradiation performance characterization was carried out on a separate test bench, with annealing occurring only passively during storage at room temperature for a few hours: a scenario representative of realistic operating conditions. No measurable performance degradation was observed; any differences between pre- \cite{Testbed, Testbed2} and post-irradiation measurements were indistinguishable from statistical variation.
Single-event upset (SEU) resilience tests under heavy-ion and neutron irradiation have not yet been performed due to resource constraints. However, no significant deviations are expected, as the design strategies employed are consistent with those used in other functionally similar ASICs that have been successfully qualified and integrated into LHC's detectors.  
Input currents down to 1~pA were generated using a calibrated source-measure unit with guarded triaxial connections (Keithley 6430 equipped with remote amplifier module). Special care was taken in the PCB design to further reduce leakage \cite{Testbed}, including the use of shielding and minimizing the distance between connectors and the ASIC to about 2~cm.

Data acquisition was carried out using a BLM crate housing a VFC acquisition board, capable of storing up to 2 Gbit raw bit-stream per measurement. The acquired data were then exported to a PC for post-processing.
Each measurement corresponds to a fixed DC input current. The equivalent noise current is evaluated as a function of acquisition time by segmenting the raw output into consecutive, non-overlapping bitstream intervals. For each segment, the average density of '1's is computed using a rectangular time window, which corresponds to a sinc-shaped frequency response. The noise is then taken as the standard deviation of these repeated measurements. The resulting input-referred noise current assumes a perfectly linear (or calibrated) transfer characteristic.

\begin{figure}
    \centering
    \includegraphics[width=1.05\linewidth]{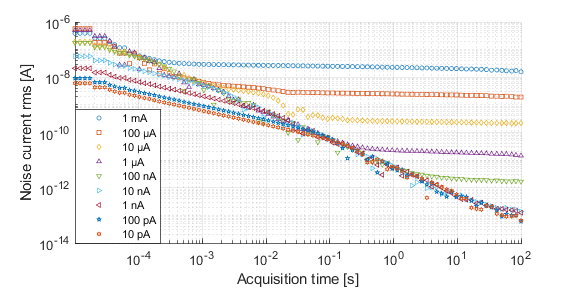}
    \caption{Equivalent input-referred noise current as a function of integration time. The noise is obtained from the standard deviation of consecutive measurements, each corresponding to the averaged bitstream over the indicated duration.}
    \label{fig:Noise_curr_meas}
\end{figure}

Figure~\ref{fig:Noise_curr_meas} reports the equivalent input-referred noise as a function of acquisition time for input currents spanning from 1~mA down to 10 pA. 
The most relevant measurement combinations are detailed in Table~\ref{tab:current_time}: sub-pA accuracy can be reached under, at least, ten seconds of integration and input currents below 10~nA. The dominant noise source shifts with input levels; as predicted by simulations, the compensation DAC limits accuracy at higher current ranges, while the operational amplifier dominates at lower ranges.
Grayed-out cells in the table indicate incompatible parameter combinations for which the noise exceeds the targeted measurement range. The last row reports the operating dynamic range corresponding to each integration window and defined as:

\begin{equation}
    DR_{oper}=20\cdot log_{10}\left( \frac{I_{in, ~max}}{i_{noise, ~min}}\right)
\end{equation}
For a 100~s integration window, this dynamic range reaches 201.9~dB.
Conversely, Table~\ref{tab:enob} reports the effective number of bits (ENOB) for each current, measured with the minimum, mid and maximum integration windows, according to:
\begin{equation}
    ENOB(I_{in},T_{acq}) = log_2\left( \frac{I_{in}}{i_{noise}(T_{acq})}\right).
\end{equation}
With a $10~\mu s$ integration window, 11~bit effective resolution is achieved only for the highest current. At 100 ms, however, 15~bit accuracy extends over three decades, from 1~mA to $10~\mu A$. For 100~s, 15~bit ENOB is exceeded for currents spanning from 1~mA down to 10~nA.

\begin{figure}
    \centering
    \includegraphics[width=\linewidth]{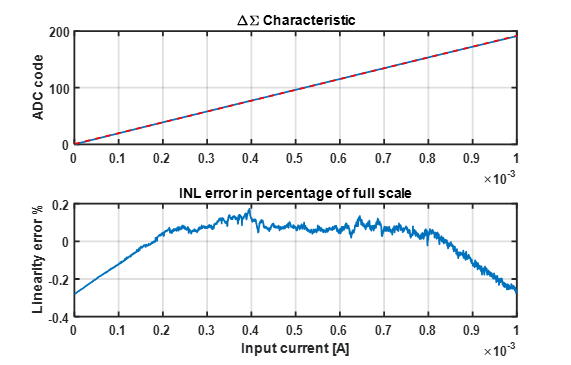}
    \caption{ADC linearity performance for a $10~\mu$s acquisition window. The top plot shows the measured transfer characteristic, while the bottom plot presents the integral nonlinearity (INL). Without calibration, the INL exhibits a maximum deviation of [+0.2\%, -0.3\%] which corresponds to [+4, -5] LSBs.}
    \label{fig:Linearity_10us}
\end{figure}

Figure~\ref{fig:Linearity_10us} presents the measured ADC transfer characteristic and integral non-linearity (INL) for the fastest acquisition setting of $10~\mu\mathrm{s}$; sampling at 20 MHz can only produce 200 distinct output codes. Across the corresponding input range of $1~\mathrm{mA}$ to $5~\mu\mathrm{A}$, the uncalibrated INL stays within $+0.2\%$ and $-0.3\%$ of full scale, complying with BLM specifications. 

\begin{figure}
    \centering
    \includegraphics[width=0.7\linewidth]{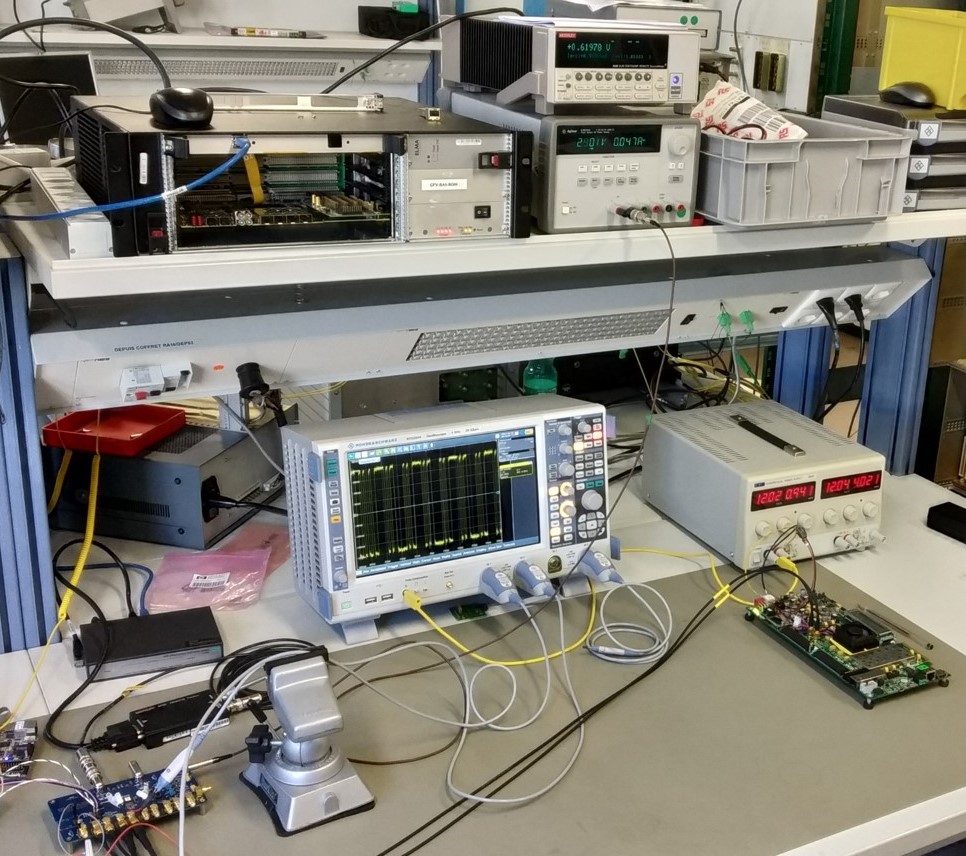}
    \caption{Photograph of the experimental measurement setup. The upper shelf hosts the BLM VFC crate, a sub-femtoampere-capable source measure unit (Keithley 6430), and a power supply. On the lower table are the test PCB, an oscilloscope, the clock generation PCB, and its power supply.}
    \label{fig:placeholder}
\end{figure}

\begin{figure}
    \centering
    \includegraphics[width=0.4\linewidth]{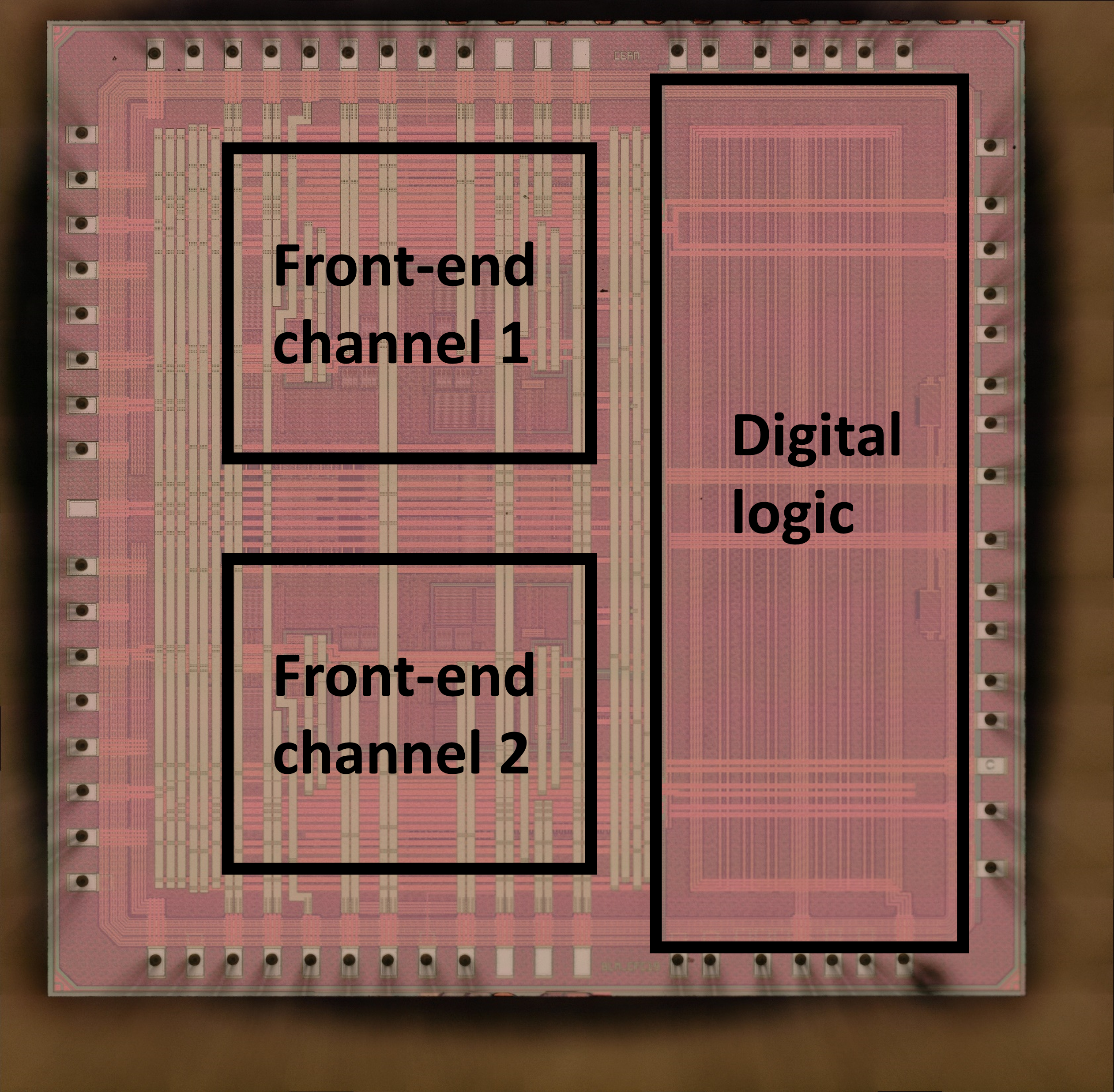}
    \caption{Micrograph of the fabricated chip. With dimensions of $3.75 \times 3.75~mm^2$ and 64 bond pads, the die integrates two analog channels on the left side and the digital control logic on the right.}
    \label{fig:chip}
\end{figure}

\begin{table*}
    \centering
    \caption{Measured RMS Noise Current vs. Input Current $I_{\text{in}}$ and Integration Time $T_{\text{acq}}$}
    \label{tab:current_time}
    \begin{tabular}{l *{8}{c}}
        \toprule
        \textbf{Input Current} & \multicolumn{8}{c}{\textbf{Integration Time} -- $T_{\text{acq}}$} \\
        \textbf{$I_{\text{in}}$} & \textbf{10~$\mu$s} & \textbf{100~$\mu$s} & \textbf{1~ms} & \textbf{10~ms} & \textbf{100~ms} & \textbf{1~s} & \textbf{10~s} & \textbf{100~s} \\
        \cmidrule(lr){2-9}
        1~mA      & 388~nA &  62~nA & 29~nA & 27~nA & 25~nA & 23~nA & 20~nA & 15~nA \\
        100~$\mu$A & 570~nA &  50~nA & 6.6~nA & 3.7~nA & 2.7~nA & 2.5~nA & 2.3~nA & 2.2~nA \\
        10~$\mu$A  & 192~nA &  56~nA & 5.8~nA & 2.2~nA & 375~pA & 256~pA & 235~pA & 217~pA \\
        1~$\mu$A   & 497~nA & 245~nA & 5.4~nA & 502~pA &  60~pA &  23~pA &  19~pA &  14~pA \\
        100~nA    & \cellcolor{lightgray}174~nA &  51~nA & 4.6~nA & 598~pA &  31~pA & 5.2~pA & 2.1~pA & 1.6~pA \\
        10~nA     & \cellcolor{lightgray}57~nA  & \cellcolor{lightgray}18~nA & 5.1~nA & 552~pA &  51~pA & 4.4~pA & 515~fA & 134~fA \\
        1~nA      & \cellcolor{lightgray}21~nA  & \cellcolor{lightgray}6.8~nA & \cellcolor{lightgray}2.1~nA & 555~pA &  46~pA & 3.9~pA & 530~fA & 123~fA \\
        100~pA    & \cellcolor{lightgray}9.2~nA & \cellcolor{lightgray}2.9~nA & \cellcolor{lightgray}922~pA & \cellcolor{lightgray}283~pA & 56~pA & 5.7~pA & 575~fA &  81~fA \\
        10~pA     & \cellcolor{lightgray}6.1~nA & \cellcolor{lightgray}2.0~nA & \cellcolor{lightgray}613~pA & \cellcolor{lightgray}191~pA & \cellcolor{lightgray}52~pA & 4.9~pA & 566~fA &  77~fA \\
        \midrule
        \multicolumn{9}{c}{\textbf{Operational Dynamic Range} -- $\mathbf{DR_{\text{oper}}}$} \\
        \quad dB  & 66.1 & 85.9 & 105.9 & 125.1 & 145.0 & 166.2 & 184.9 & 202.3 \\
        \quad bits & 11.0 & 14.3 & 17.6 & 20.8 & 24.1 & 27.6 & 30.7 & 33.6 \\
        \bottomrule
    \end{tabular}
    
    \smallskip
    \footnotesize
    \centering
    Gray cells indicate configurations where the measured noise exceeds the nominal input current and are therefore not included in the $DR_{oper}$ estimation.
\end{table*}

\begin{table}
    \centering
    \caption{ENOB vs integration times and input currents}
    \label{tab:enob}
    \begin{tabular}{l *{3}{c} }
    
        \toprule 
        {Input Current} & \multicolumn{3}{c}{$ENOB(T_{acq})$} \\
        \cmidrule(lr){2-4} % Changed from {2-9} to {2-4}
        {\(I_{\text{in}}\)} & {$T_{\text{int}} = 10~\mu\text{s}$} & {$T_{\text{int}} = 100~\text{ms}$} & {$T_{\text{int}} = 100~\text{s}$} \\ 
        \midrule
        1~\(m\)A  &  11.33 & 15.28 & 15.96\\
        100~\(\mu\)A  & 7.45 & 15.16 & 15.78\\
        10~\(\mu\)A & 5.70 & 14.70 & 15.49\\
        1~\(\mu\)A &--& 14.02 & 16.12\\
        100~\(n\)A &--& 11.64& 15.88\\
        10~\(n\)A &--& 7.61 & 16.18\\
        1~\(n\)A &--& 4.43 & 12.99\\
        100~\(p\)A &--&--& 10.26\\
        10~\(p\)A  &--&--& 7.02\\
        \bottomrule
    \end{tabular}
\end{table}

The power consumption of the 2-channel ASIC is below 25~mW from a single 1.2 V supply.

\section{Conclusion}
This paper has presented a radiation-tolerant current digitizer ASIC designed for beam loss monitoring applications in the HL-LHC environment. The architecture is derived directly from stringent system-level requirements, including a wide input dynamic range [1~mA,~1~pA], short integration windows down to 10~\textmu s, deterministic overload recovery, and tolerance to total ionizing doses exceeding 100~Mrad.

A first-order delta-sigma topology was selected to ensure robust bounded behavior within its operational range, predictable saturation behavior, and robustness under large transient inputs, as required in machine protection systems. 
The achievable operating dynamic range, which is a function of the input current, scales with the acquisition time. This property allows the converter to deliver high resolution for quasi-static signals while maintaining the fast response required for transient events. In 10~\textmu s mode, the system achieves 11~bit resolution, while extended acquisition enables significantly higher effective resolution.

The circuit implementation in a 130~nm CMOS process incorporates radiation-hardening techniques at both architectural and layout levels. Experimental results demonstrate correct functionality and no measurable performance degradation after exposure to 100~Mrad under worst-case bias conditions. Noise measurements confirm pA-level resolution, and linearity results validate correct operation across the full input range.

The proposed approach demonstrates that wide dynamic range, deterministic behavior, and high radiation tolerance can be simultaneously achieved within a compact monolithic solution. Although developed for the LHC Beam Loss Monitoring system, the architecture is applicable to other high-radiation and wide-dynamic-range current sensing applications in high-energy physics and related fields

\section*{Acknowledgments}
The author would like to thank the CERN EP-ESE Group and the BE-BI Group for their contribution to the development of this work. Special thanks are due to Jan Kaplon for his invaluable support in analog ASIC design and exquisite mentorship, to Pedro Leitao for the design of the digital part, to Giulio Borghello for assistance with irradiation tests, to Francesco Martina for the PCB design and to Christos Zamantzas for providing access anf thechnical support to the BL testing laboratory.

\end{document}